\documentclass[12pt]{article}
\usepackage{epsfig}
\usepackage{amsmath}
\usepackage{hhline}
\usepackage{amssymb}
\usepackage{times}

\newlength{\dinwidth}
\newlength{\dinmargin}
\begin{document}
\newcommand{\pom}{{I\!\!P}}
\newcommand{\reg}{{I\!\!R}}
\newcommand{\slowpi}{\pi_{\mathit{slow}}}
\newcommand{\fiidiii}{F_2^{D(3)}}
\newcommand{\fiidiiiarg}{\fiidiii\,(\beta,\,Q^2,\,x)}
\newcommand{\n}{1.19\pm 0.06 (stat.) \pm0.07 (syst.)}
\newcommand{\nz}{1.30\pm 0.08 (stat.)^{+0.08}_{-0.14} (syst.)}
\newcommand{\fiidiiiful}{F_2^{D(4)}\,(\beta,\,Q^2,\,x,\,t)}
\newcommand{\fiipom}{\tilde F_2^D}
\newcommand{\ALPHA}{1.10\pm0.03 (stat.) \pm0.04 (syst.)}
\newcommand{\ALPHAZ}{1.15\pm0.04 (stat.)^{+0.04}_{-0.07} (syst.)}
\newcommand{\fiipomarg}{\fiipom\,(\beta,\,Q^2)}
\newcommand{\pomflux}{f_{\pom / p}}
\newcommand{\nxpom}{1.19\pm 0.06 (stat.) \pm0.07 (syst.)}
\newcommand {\gapprox}
   {\raisebox{-0.7ex}{$\stackrel {\textstyle>}{\sim}$}}
\newcommand {\lapprox}
   {\raisebox{-0.7ex}{$\stackrel {\textstyle<}{\sim}$}}
\def\gsim{\,\lower.25ex\hbox{$\scriptstyle\sim$}\kern-1.30ex%
\raise 0.55ex\hbox{$\scriptstyle >$}\,}
\def\lsim{\,\lower.25ex\hbox{$\scriptstyle\sim$}\kern-1.30ex%
\raise 0.55ex\hbox{$\scriptstyle <$}\,}
\newcommand{\pomfluxarg}{f_{\pom / p}\,(x_\pom)}
\newcommand{\dsf}{\mbox{$F_2^{D(3)}$}}
\newcommand{\dsfva}{\mbox{$F_2^{D(3)}(\beta,Q^2,x_{I\!\!P})$}}
\newcommand{\dsfvb}{\mbox{$F_2^{D(3)}(\beta,Q^2,x)$}}
\newcommand{\dsfpom}{$F_2^{I\!\!P}$}
\newcommand{\gap}{\stackrel{>}{\sim}}
\newcommand{\lap}{\stackrel{<}{\sim}}
\newcommand{\fem}{$F_2^{em}$}
\newcommand{\tsnmp}{$\tilde{\sigma}_{NC}(e^{\mp})$}
\newcommand{\tsnm}{$\tilde{\sigma}_{NC}(e^-)$}
\newcommand{\tsnp}{$\tilde{\sigma}_{NC}(e^+)$}
\newcommand{\st}{$\star$}
\newcommand{\sst}{$\star \star$}
\newcommand{\ssst}{$\star \star \star$}
\newcommand{\sssst}{$\star \star \star \star$}
\newcommand{\tw}{\theta_W}
\newcommand{\sw}{\sin{\theta_W}}
\newcommand{\cw}{\cos{\theta_W}}
\newcommand{\sww}{\sin^2{\theta_W}}
\newcommand{\cww}{\cos^2{\theta_W}}
\newcommand{\trm}{m_{\perp}}
\newcommand{\trp}{p_{\perp}}
\newcommand{\trmm}{m_{\perp}^2}
\newcommand{\trpp}{p_{\perp}^2}
\newcommand{\alp}{\alpha_s}

\newcommand{\alps}{\alpha_s}
\newcommand{\sqrts}{$\sqrt{s}$}
\newcommand{\LO}{$O(\alpha_s^0)$}
\newcommand{\Oa}{$O(\alpha_s)$}
\newcommand{\Oaa}{$O(\alpha_s^2)$}
\newcommand{\PT}{p_{\perp}}
\newcommand{\JPSI}{J/\psi}
\newcommand{\sh}{\hat{s}}
\newcommand{\uh}{\hat{u}}
\newcommand{\MP}{m_{J/\psi}}
\newcommand{\PO}{I\!\!P}
\newcommand{\xbj}{x}
\newcommand{\xpom}{x_{\PO}}
\newcommand{\ttbs}{\char'134}
\newcommand{\xpomlo}{3\times10^{-4}}  
\newcommand{\xpomup}{0.05}  
\newcommand{\dgr}{^\circ}
\newcommand{\pbarnt}{\,\mbox{{\rm pb$^{-1}$}}}
\newcommand{\gev}{\,\mbox{GeV}}
\newcommand{\WBoson}{\mbox{$W$}}
\newcommand{\fbarn}{\,\mbox{{\rm fb}}}
\newcommand{\fbarnt}{\,\mbox{{\rm fb$^{-1}$}}}
%
%
\newcommand{\qsq}{\ensuremath{Q^2} }
\newcommand{\gevsq}{\ensuremath{\mathrm{GeV}^2} }
\newcommand{\et}{\ensuremath{E_t^*} }
\newcommand{\rap}{\ensuremath{\eta^*} }
\newcommand{\gp}{\ensuremath{\gamma^*}p }
\newcommand{\dsiget}{\ensuremath{{\rm d}\sigma_{ep}/{\rm d}E_t^*} }
\newcommand{\dsigrap}{\ensuremath{{\rm d}\sigma_{ep}/{\rm d}\eta^*} }
\def\Journal#1#2#3#4{{#1} {\bf #2} (#3) #4}
\def\NCA{\em Nuovo Cimento}
\def\NIM{\em Nucl. Instrum. Methods}
\def\NIMA{{\em Nucl. Instrum. Methods} {\bf A}}
\def\NPB{{\em Nucl. Phys.}   {\bf B}}
\def\PLB{{\em Phys. Lett.}   {\bf B}}
\def\PRL{\em Phys. Rev. Lett.}
\def\PRD{{\em Phys. Rev.}    {\bf D}}
\def\ZPC{{\em Z. Phys.}      {\bf C}}
\def\EJC{{\em Eur. Phys. J.} {\bf C}}
\def\CPC{\em Comp. Phys. Commun.}

\newcommand{\dps}{\displaystyle}
\begin{titlepage}

\noindent

\begin{flushleft}
DESY-00-143 \hfill ISSN0418-9833 \\
September 2000
\end{flushleft}

\vspace{2cm}

\begin{center}
\begin{Large}

\boldmath
{\bf Dijet Production in Charged and Neutral Current \\
$e^{+}p$ Interactions at High $Q^2$}
\unboldmath
 
\vspace{2cm}

H1 Collaboration

\end{Large}
\end{center}

\vspace{2cm}

\begin{abstract}
\noindent
Jet production in charged and neutral current events in the 
kinematic range of $Q^2$ from 640 to 35\,000~GeV$^2$ 
is studied in deep-inelastic positron-proton scattering at HERA. The 
measured rate of multi-jet events and distributions of jet polar angle, 
transverse energy, dijet mass, and other dijet variables are presented. 
Using parton densities derived from inclusive DIS cross sections, 
perturbative QCD calculations in NLO are found to give a consistent 
description of both the neutral and charged current dijet production. 
A direct, model independent comparison of the jet distributions in 
charged and neutral current events
confirms that the QCD dynamics of the hadronic final state is 
independent of the underlying electroweak scattering process.
\end{abstract}

\centerline{\it submitted to Eur. Phys. J. C }

\end{titlepage}

\clearpage
\newpage
 
 \begin{flushleft}
 C.~Adloff$^{33}$,                
 V.~Andreev$^{24}$,               
 B.~Andrieu$^{27}$,               
 V.~Arkadov$^{35}$,               
 A.~Astvatsatourov$^{35}$,        
 I.~Ayyaz$^{28}$,                 
 A.~Babaev$^{23}$,                
 J.~B\"ahr$^{35}$,                
 P.~Baranov$^{24}$,               
 E.~Barrelet$^{28}$
% ...
 \end{flushleft}
 
 \begin{flushleft} 
% ...
$ ^2$ III. Physikalisches Institut der RWTH, Aachen, Germany$^a$ \\
% ...
$ ^{23}$ Institute for Theoretical and Experimental Physics,
          Moscow, Russia \\
 $ ^{24}$ Lebedev Physical Institute, Moscow, Russia$^{e,h}$ \\
% ...
$ ^{27}$ LPNHE, \'{E}cole Polytechnique, IN2P3-CNRS, Palaiseau, France \\
 $ ^{28}$ LPNHE, Universit\'{e}s Paris VI and VII, IN2P3-CNRS,
          Paris, France \\
% ...
$ ^{33}$ Fachbereich Physik, Bergische Universit\"at Gesamthochschule
          Wuppertal, Wuppertal, Germany$^a$ \\
% ...
$ ^{35}$ DESY, Zeuthen, Germany$^a$ \\
% ...
$ ^a$ Supported by the Bundesministerium f\"ur Bildung, Wissenschaft,
        Forschung und Technologie, FRG,
        under contract numbers 7AC17P, 7AC47P, 7DO55P, 7HH17I, 7HH27P,
        7HD17P, 7HD27P, 7KI17I, 6MP17I and 7WT87P \\
% ...
$ ^e$ Supported by the Deutsche Forschungsgemeinschaft \\
% ...
$ ^h$ Supported by Russian Foundation for Basic Research 
       grant no. 96-02-00019 \\
% ...
 \end{flushleft}
 
\clearpage
\newpage

\section{Introduction}
Deep-inelastic scattering (DIS) at the electron-proton collider HERA offers 
unique possibilities to reveal the partonic structure of matter. At  very 
high four-momentum transfer squared $-Q^2$ the exchange of all the 
electroweak gauge bosons (photon, $Z^0$ and $W^{\pm}$) becomes important 
allowing the standard model of electroweak and strong interactions to be 
tested at distances as small as 10$^{-18}$m. 
The inclusive DIS cross sections of neutral current (NC) $ep\rightarrow eX$ 
and charged 
current (CC) $ep\rightarrow \nu X$ interactions have been measured 
\cite{highQ2,highQ2Zeus} and are 
well described by the standard model. 
In this analysis we complement these results by the first detailed 
investigation of dijet structures in both NC and CC processes.

\begin{figure}[t]
\vspace*{3.cm}
\begin{center}
\epsfig{file=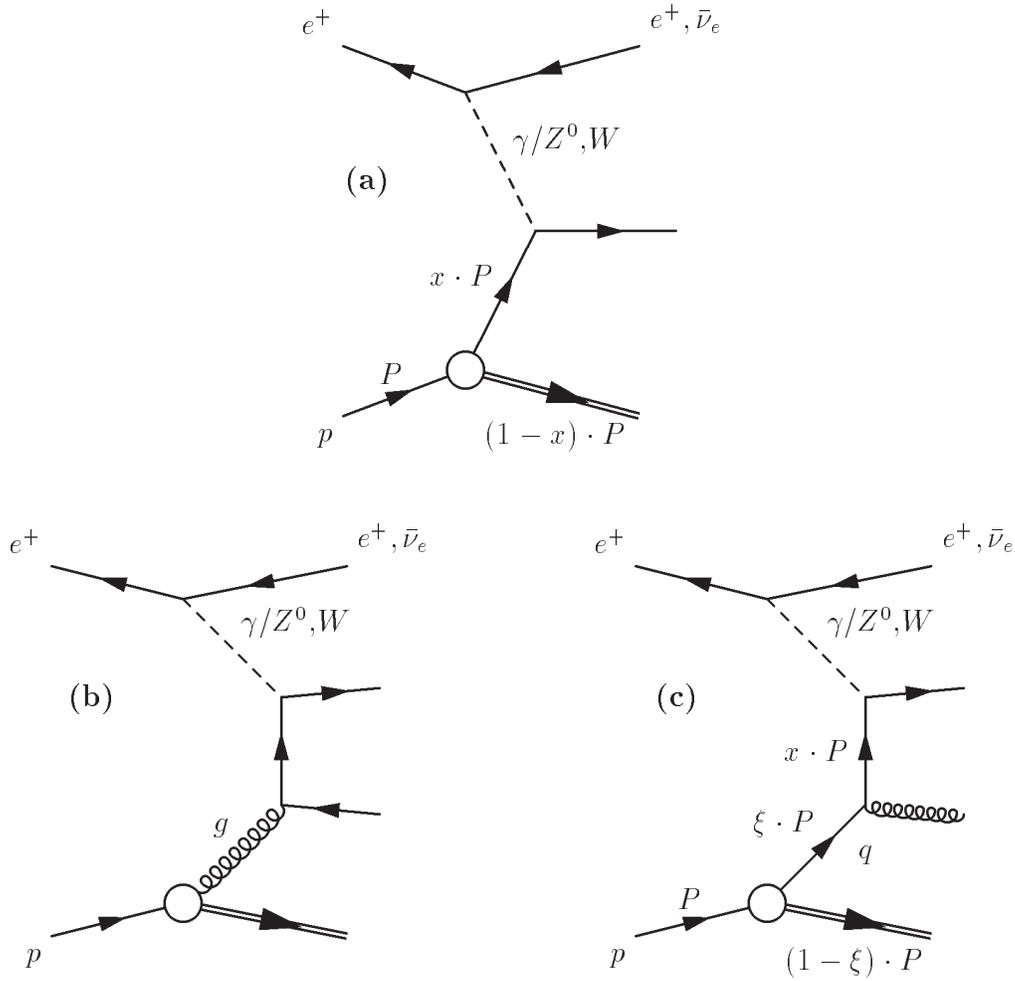, bbllx=100pt,bblly=200pt,bburx=650pt,bbury=700pt,
width=19.cm}
\end{center}
\caption{Feynman graphs for DIS in lowest order (a), and selected 
leading-order diagrams contributing to dijet production: 
Boson--Gluon--Fusion (b) and QCD--Compton scattering (c). The variables 
$x$ and $\xi$ denote the fraction of the proton's momentum $P$ carried 
by the scattered parton.}
\label{bgfqcdc}
\end{figure}

Within the Quark--Parton--Model DIS gives rise to events with (1+1) jets, 
where one jet originates from a quark struck out of the proton 
and a second jet is due to the proton remnant (denoted `+1'). Events with 
(2+1) jets, referred to as dijet events, are predicted by Quantum 
Chromodynamics (QCD) due to contributions in $O(\alpha_s)$, namely 
QCD--Compton scattering $eq \rightarrow eq g$ 
and Boson--Gluon--Fusion $eg \rightarrow eq\bar{q}$ 
as illustrated in Figure~\ref{bgfqcdc}. In CC interactions several events 
with multijet structures have been identified \cite{ZEUS.CC.R2} and the jet 
shape has been measured \cite{ZEUS.CC.jetshap}. However, due to the 
relatively small number of CC events observed so far at HERA, the structure 
of the hadronic final state has not yet been studied in detail. 
In NC interactions clear multi-jet structures have been established
\cite{multijet} and have been used to test QCD \cite{Jet.NC}. 
Previous analyses of dijet production in NC processes, however, did not yet 
extend to very high values of $Q^2$.

In the present paper a dijet analysis of a sample of 460 CC 
events and approximately 8\,600 NC events with $Q^2$ in the range of 
640 to 35\,000~GeV$^2$ 
is performed. 
Various dijet distributions are compared with the 
predictions of QCD Monte Carlo models
and with perturbative QCD calculations in next-to-leading order (NLO). 
In addition, the jet distributions of the CC and the NC events are  
compared directly, in order to test the 
hypothesis that QCD radiation proceeds independently of the underlying 
electroweak scattering process.

\section{Detector description and data selection}

This analysis is based on the data sample recorded with the H1 detector 
in the data taking periods 1994--1997  at HERA. In this period HERA was 
operated with positron and proton beams of 27.5 and 820 GeV respectively, 
corresponding to a centre--of--mass energy of $\sqrt{s} \approx 300$ 
GeV. The collected integrated luminosity for this analysis is 35.6 pb$^{-1}$. 

\subsection{Detector and trigger}

The components of the H1 detector \cite{H1-detector} most relevant for this 
analysis are the central tracking system, the liquid argon calorimeter and 
the instrumented iron return yoke. 

The central tracking system consists of two concentric drift chambers 
covering a polar angular range\footnote{The forward direction and the 
positive $z$-axis are defined as the proton beam direction. The origin of 
coordinates is the nominal $ep$ interaction point.} of 
15$^{\circ}$ to 165$^{\circ}$. Two 
polygonal drift chambers with wires perpendicular to the beam axis improve 
the determination of the $z$ coordinate of the measured tracks. 
The central tracking system is surrounded by a liquid argon sampling 
calorimeter covering a 
polar angle range of 4$^{\circ}<\theta<154^{\circ}$. 
The electromagnetic and hadronic sections of 
the liquid argon calorimeter correspond in total to a depth of 4.5 to 8 
interaction lengths. The energy resolution of the liquid argon calorimeter 
for electrons and hadrons was determined  in test beam 
measurements to be 
$\sigma/E = 12\%/\sqrt{E({\rm GeV})} \oplus 1\%$ and 
$\sigma/E = 50\%/\sqrt{E({\rm GeV})} \oplus 2\%$, respectively 
\cite{lar-res}. The systematic uncertainty of the 
electromagnetic energy scale is determined to be 0.7\% for the majority of 
the selected events and increases to 3\% at the highest $Q^2$ \cite{highQ2}. 
The uncertainty on the hadronic energy scale of the liquid argon calorimeter 
is 4\%.

Outside the calorimeters a large superconducting solenoid provides 
a magnetic field of 1.15~Tesla. The instrumented iron return yoke 
identifies energetic muons and detects leakage of hadronic showers.

The trigger conditions for CC events are based on the reconstruction 
of a large missing transverse momentum in the trigger sums of the liquid argon 
calorimeter \cite{lar-towers}. NC events are triggered on the basis of 
a localized high energy deposit in the electromagnetic part of the calorimeter 
\cite{H1-detector}. 

\subsection{Event selection}

\noindent
{\bf Selection of CC events}
\newline\noindent
The selection of CC events is similar to those of \cite{highQ2,CC-sel}. It is 
based on the observation of a large imbalance in transverse momentum due to 
the antineutrino escaping direct detection. The transverse momentum 
$P_T^{had}$, reconstructed with the liquid argon calorimeter and
the instrumented iron, is required to exceed 25 GeV.
No scattered positron must be found in order to reject neutral current 
events. The $z$ coordinate of the primary event vertex $z_{vtx}$ has to be 
within a distance of $\pm$35 cm from the nominal $ep$ collision point. The 
inelasticity $y_{had}=\sum_h E_h (1-\cos\theta_h)/2E_e$, calculated from 
the energy depositions in the calorimeters and the energy of the positron 
beam $E_e$, must be in the range $0.03 < y_{had} < 0.85$. 
The kinematic selection criteria imply a minimum virtuality $Q^2$ of the 
exchanged boson of 640 GeV$^2$. 

Background events due to cosmic muons, beam-halo muons and beam-gas 
interactions are removed by further requirements on the 
event topology and timing \cite{CC-sel}. Furthermore a visual scan of 
the remaining events is performed.

The final event sample consists of 460 CC events.  
The background from photoproduction events is less than 2\%. It is estimated 
from Monte Carlo simulations and from data events where the scattered 
electron is detected at very small scattering angles. The number of 
background events from other sources is negligible.

The trigger efficiency for events with $P_T^{had} > 25$ GeV 
has been determined  as a function of the kinematic variables and the 
jet variables studied using NC events where the information of the 
scattered positron \cite{CC-sel} is discarded. The average trigger efficiency 
is $\approx 95\%$. 
It is corrected for in all measured distributions.

\noindent
{\bf Selection of NC events}
\newline\noindent
The kinematic selection criteria of the NC events correspond to those of 
the CC events. The NC selection requires the identification of the scattered 
positron. Fiducial cuts are applied to the impact position of the scattered 
positron in the liquid 
argon calorimeter in order to avoid inhomogeneities at the 
boundary of detector modules 
\cite{highQ2}. The kinematic selection is based on the variables $p_T^e$ 
and $y_e$ reconstructed from the scattered positron momentum with the 
exception of the requirement $y_{had} > 0.03$. The summed energy $E$ and 
longitudinal momentum components $P_z$ of all reconstructed detector objects 
(see section 4.1) must fulfil $E-P_{z} > 35$ GeV to suppress QED radiative 
events. The $z$ coordinate of the primary event vertex is required to be in 
the same range as for the CC events. The NC sample consists of 
$\approx$ 8\,600 events with a negligible number of background events from 
photoproduction. The trigger efficiency of the NC events is $\approx 99\%$.

\noindent

The distributions of the reconstructed kinematic variables $P_T$, $Q^2$, the 
Bjorken scaling variable $x$ and $y$ for the CC and the NC event samples are 
shown in Figure~\ref{kin.CC.NC.raw}. 
A good description of the data by the Monte Carlo model ARIADNE (see next 
section) 
combined with the H1 detector simulation is observed.
Note that the distributions are normalized to the total number of 
CC or NC events $N_{DIS}$ respectively. The differences between the CC and 
NC distributions \cite{highQ2} are due to the different couplings and 
propagators of the bosons in CC and NC interactions.

The main selection criteria for the CC and NC events are summarized in Table 
\ref{tabcuts}.

\begin{table}[h!]
\begin{center}
\begin{tabular}{|c|c|} \hline
         CC   & NC  \\ \hline\hline 
no $e^+$ found & $e^+$ found \\
$P_T^{had} > 25$ GeV & $p_T^e > 25$ GeV \\
$0.03 < y_{had} < 0.85 $  & $0.03 < y_{had}$, $y_e < 0.85$ \\ \hline    
 --                   &  $E-P_z > 35$ GeV    \\ 
$|z_{vtx}| < 35$ cm &  $|z_{vtx}| < 35$ cm  \\ \hline
\end{tabular}
\caption{Selection criteria for the CC and NC DIS event samples}
\label{tabcuts}
\end{center}
\end{table}

\section{QCD Monte Carlo models and QCD NLO programs}

\subsection{QCD Monte Carlo models}

Four different QCD Monte Carlo models are used in this analysis:
ARIADNE 4.10 \cite{Ariadne}; HERWIG 5.9  \cite{herwig},  
LEPTO 6.5.2$\beta$ \cite{Lepto} and  RAPGAP 2.08/06 \cite{Rapgap}. All models 
use the LO matrix elements for QCD-Compton and Boson-Gluon-Fusion. 
ARIADNE implements higher order QCD processes with radiating colour dipoles 
\cite{CDM}, HERWIG, LEPTO and RAPGAP 
use initial and final state parton showers instead \cite{PS}. In the 
context of this analysis LEPTO and RAPGAP are similar. They differ in the way 
the divergences of the LO matrix element are regulated. Fragmentation 
of partons into hadrons is modelled with the Lund string model \cite {string} 
in ARIADNE, LEPTO and RAPGAP, and with the cluster model \cite{clufrag} in 
HERWIG. 

The latest versions of the models as described in \cite{HERA-MC} are used.
The LEPTO version used contains a refinement of soft-colour interactions, 
the generalised area law model \cite{GAL}. LEPTO has been tuned to describe jet distributions at HERA and the 
corresponding values of the model parameters are taken here. 
In HERWIG, we use the leading order and not the next-to-leading order formula 
for $\alpha_s$ as proposed in \cite{HERA-MC}. The parton density 
functions CTEQ4L \cite{CTEQ4L} are taken. 

ARIADNE and LEPTO are incorporated into DJANGO \cite{Django}, version 6.2, 
which simulates the effects of QED radiation.

\subsection{QCD NLO programs}

Four programs MEPJET \cite{Mepjet}, DISENT \cite{Disent}, DISASTER++ 
\cite{Disaster} and JETVIP \cite{Jetvip} are available for perturbative 
QCD calculations of jet cross sections in NLO. Currently MEPJET is the 
only NLO program that considers $W$ or $Z^0$ exchange. MEPJET is thus used 
to calculate the jet distributions in CC processes. The NC jet cross sections 
are calculated with DISENT following the recommendations in \cite{DGWG20}. We 
use the parton density functions determined by the H1 Collaboration 
\cite{highQ2} and choose  $Q^2$ as the renormalization and factorization 
scales, $\mu_R^2$ and $\mu_F^2$. 

We compared the predictions of MEPJET and DISENT 
for the jet distributions presented below. In leading order, we 
find agreement within a fraction of a per cent. In NLO, MEPJET is 
systematically lower than DISENT by $\approx 10 \%$ confirming the 
results of \cite{WG20}. Note that the comparison of various jet cross 
sections calculated with DISENT and DISASTER++ showed good agreement 
\cite{WG20}. The differences between DISENT and DISASTER++ observed for 
extreme values of event shape variables \cite{WG20EV,Salam} are not relevant 
to this analysis. Currently it is unknown if the 
inconsistencies between the NLO programs observed in NC influence the CC 
predictions as well. 

DISENT does not consider $Z^0$ exchange, which for $Q^2 >$ 5\,000 GeV$^2$  
reduces the inclusive $e^+p$ NC cross section by less than 5\% on average 
compared with purely electromagnetic exchange. Since the dijet cross sections 
are also reduced, the effect on jet distributions normalized to the number of 
DIS events is small. Correction factors 
were calculated using ARIADNE 4.10 and are applied to any DISENT prediction 
at $Q^2 >$ 5\,000 GeV$^2$.

In order to compare the perturbative QCD predictions to the data, 
bin-by-bin hadronization corrections are determined using the QCD models 
ARIADNE and HERWIG. The average correction factors from the two models are 
applied to the NLO distributions. The maximum deviation between 
the average correction factor and the correction factor for either model 
alone is taken as hadronization uncertainty.

\section{Definition of jet observables}

\subsection{Jet algorithm}

Jets are reconstructed with a modified version of the Durham jet algorithm 
which was originally introduced in e$^{+}$e$^{-}$ annihilation experiments 
\cite{Durham}. The algorithm is applied in the laboratory frame. 
It is modified for application in DIS in two respects: 
a missing-momentum four--vector is introduced which is treated as an 
additional object by the jet algorithm to account for the momentum carried 
by the proton remnant escaping through the beam pipe;
in NC events the scattered positron is removed from the final state objects 
and is only used to determine the missing-momentum vector. In CC events 
this is achieved by first reconstructing the neutrino from the hadronic final 
state, exploiting energy and momentum conservation.

The jet algorithm calculates the quantity 
$k_{T,\;ij}^{2}=2 \min[E_{i}^2, E_{j}^2]\,(1-\cos\theta_{ij})$ 
of pairs of objects or `proto' jets $i$,~$j$. Here $E_{i}$ and $E_{j}$ are the 
energies of the objects $i$ and $j$, and $\theta_{ij}$ is the angle between 
them. The jet algorithm combines the pair of objects $i$, $j$ with the 
minimum $k_{T,\;ij}^{2}$ to be 
a `proto' jet by adding their four--momenta $p_{i}$ and $p_{j}$. This prescription is 
repeated iteratively for the remaining objects until exactly (2+1) jets remain. 
At this stage, every event is treated as a dijet event by definition. Pronounced dijet structures 
are then selected by imposing a lower limit on $y_2$, defined as the minimum $k_{T,ij}^2/W^2$ of any 
combination of the (2+1) jets. Here, $W$ is the invariant mass of the hadronic final state. 
It is calculated from all objects entering the jet algorithm.

In order to determine the fraction of events with say (1+1) or (3+1) like jet
structures it is more convenient to run the algorithm with a fixed jet resolution
parameter $y_{cut}$. In this case the iterative clustering procedure ends, when the
$k_{T,ij}^2/W^2$ of any pair of objects or proto jets is larger than a given value 
$y_{cut}$. Thus the number of jets reconstructed varies from event to event. 

The algorithm is  applied to the tracks reconstructed in the central tracking 
chambers and the energy depositions (clusters) in the liquid argon 
calorimeter. For tracks and calorimeter clusters that can be matched, 
the energy is determined from either the calorimeter alone or from a 
combination of track and cluster energy as described in \cite{highQ2}. The 
polar angle of each detector object, 
track or cluster, is required to exceed $7^{\circ}$ in order to avoid the 
region close to the edge of the calorimeter. This improves the resolution 
of the reconstructed jet quantities. 

The same definitions of the jets are used for the analysis of the data 
and the Monte Carlo events after detector simulation. In events simulated 
at the hadron or parton level and in the perturbative QCD calculations, 
the jet algorithm is applied to hadron or parton four-momenta respectively. 
The polar angle cut of 7$^{\circ}$, which is applied for detector objects, 
is also applied for hadrons and partons. 

In the calculation of $y_2$ the effects of the hadronic energy scale 
uncertainty largely cancel due to the method chosen to reconstruct $W$. 
The choice of a jet algorithm working in the laboratory frame leads to 
reduced experimental errors since a boost into another frame is avoided. This 
is relevant for the CC events where the resolution of the 
kinematic variables is worse than in NC events. 



In the present analysis the jet polar angles must fulfil $10^{\circ}
<\theta_{jet} < 140^{\circ}$.  This restricts the jets to the
acceptance of the liquid argon calorimeter.

\subsection{Jet observables}

The rate of events with jet multiplicity $i$ is defined as 
$R_i(Q^2) \equiv N_i(Q^2)/N_{DIS}(Q^2)$, where $N_{i=1,2,3}$ is the number of
events with one, two or three jets, not counting the proton remnant jet. 
$N_{DIS}$ is the number of selected deep-inelastic events. The value of 
$y_{cut}$ is taken as 0.002.

The dijet sample is correspondingly defined by the requirement $y_2>0.002$. 
Large values of $y_2$ correspond  to events with (2+1) jets that are clearly 
separated and indicative of hard QCD radiation. Small values of $y_2$ are 
typical for events which intuitively may be considered as (1+1) jet events. 
In Figure \ref{events} two CC events with very different values 
of $y_2$ are displayed for illustration. Note that, with this definition of 
the dijet sample, the events contributing to the one-jet rate as introduced 
above are eliminated, and the few three-jet events are now treated as dijet 
events. 

The distributions of the dijet variables $y_2$, $m_{12}$, $z_p$, $x_p$, the 
polar angle $\theta_{fwd}$ and the transverse energy $E_{T,fwd}$ of the most 
forward (non--remnant) jet are studied. $m_{12}$ is the invariant mass of the 
two non-remnant jets. The variables $z_p$ and $x_p$ are defined by 

$$ z_p\equiv 
 {\displaystyle\min_{i = 1,2}  [E_{i}\,(1 - \cos\theta_{i})]}/
{\sum_{i = 1,2} \,\displaystyle E_{i}\,(1 - \cos\theta_{i})} \;\;\;
{\rm and}\;\;\; x_p \equiv \frac{\displaystyle Q^2}
                                    {\displaystyle Q^2 + m_{12}^2}$$

\noindent
where $E_{i}$ and $\theta_{i}$ are the energies and polar angles of the two 
(non--remnant) jets remaining after the clustering of the jet algorithm.
The variable $z_p$ corresponds to 
$1/2\, {\displaystyle\min_{i = 1,2}}(1-\cos\theta_i^*)$ 
where $\theta^*$ is the polar angle of the parton $i$ in the 
centre-of-mass system of the virtual boson and the incoming parton. In 
leading order QCD $x_p$ is equal to the ratio $x/\xi$ where $\xi$ is the 
fraction of the proton's four momentum carried by the parton entering in the 
hard scattering process (see Figure \ref{bgfqcdc}). In the limit where one 
jet is absorbed into the remnant jet $z_p$ approaches 0. In the other limit, 
where the two (non--remnant) jets become one jet, $m_{12}$ approaches 0 and 
$x_p$ approaches 1.

\section{Correction of the data}

The data are corrected for the effects of detector acceptance and resolution, 
and of QED radiation using the QCD models ARIADNE and LEPTO. For the 
correction of CC events, the number of events simulated for either model is 
approximately 150 times larger than that of the experimental data. The number 
of simulated NC events is at least six times 
larger than that of the data. The same event cuts and track/cluster 
selection criteria are applied to the simulated events and to the data. 

\noindent
{\bf Correction of detector effects}\newline
The measured jet distributions are corrected for detector effects with 
bin-by-bin correction factors. 
The purity, defined as the number of simulated events which originate in a 
bin and are reconstructed in it, normalized by the number of reconstructed 
events in that bin, is on average 60\% for both CC and NC distributions. 
The purities estimated with LEPTO and ARIADNE are very similar. 

The stability of the results was tested by correcting the jet 
distributions reconstructed from simulated LEPTO events with the correction 
factors derived with ARIADNE. The agreement of the corrected jet 
distributions with the `true' LEPTO jet distributions is good. Deviations are 
typically of a few percent. The largest deviations of $10-15\%$ are seen in 
the $z_p$ distribution. These effects are considered in the model uncertainty 
discussed below. 

\noindent
{\bf Correction of QED radiative effects}\newline
The effects of QED radiation are considered by correction factors also.
These factors are obtained from the ratio of the Monte Carlo distributions 
generated with and without inclusion of QED effects. 
The size of the corrections is $\approx 5\%$ for both CC and NC 
distributions. 

The combined detector and QED radiation correction factors from LEPTO and 
ARIADNE are averaged, and the resulting mean correction factors are 
used to correct the jet distributions.

\section{Determination of systematic errors}

The major sources of systematic errors are the model dependence of 
the detector corrections and the uncertainties of the electromagnetic and 
the hadronic energy 
scales of the liquid argon calorimeter. The total error of the majority of 
the CC data points is dominated by the statistical errors whereas the 
statistical and systematic errors are roughly of the same size for the NC 
data.

\noindent
{\bf Model dependence of correction factors}\newline
The difference between the (average) corrected distributions and the 
distributions corrected with either model alone is taken as the error. 
The error is on average $\approx \pm 3\%$ for both the CC and NC 
distributions. 

\noindent
{\bf Electron energy calibration}\newline 
The energy scale of electrons measured in the liquid argon calorimeter is 
known to 0.7\% in the angular region where most events are situated. 
The effect on the jet distributions of the NC event sample is
generally smaller than one per cent.

\noindent
{\bf Hadronic energy calibration}\newline
In order to estimate the effect of the hadronic energy scale uncertainty
on the measured jet distributions, the analysis is repeated with the 
hadronic cluster energies shifted by $\pm 4\%$. The size of the corresponding 
changes  depends considerably on the observable studied. For the CC events,  
the largest variation of $\approx 11\%$ is observed for the $m_{12}$ 
distribution, the smallest variations of $\approx 3\%$ are observed for the 
$y_2$, $z_p$ and $\theta_{fwd}$ distributions. Similar variations are 
observed in the NC jet distributions. 
A variation of the track momentum by $\pm 3\%$ has negligible effect on the 
jet distributions.

\section{Results}

\subsection{Jet event rates}

\begin{figure}[h!]
\vspace*{-2.cm}
\begin{center}
\epsfig{file=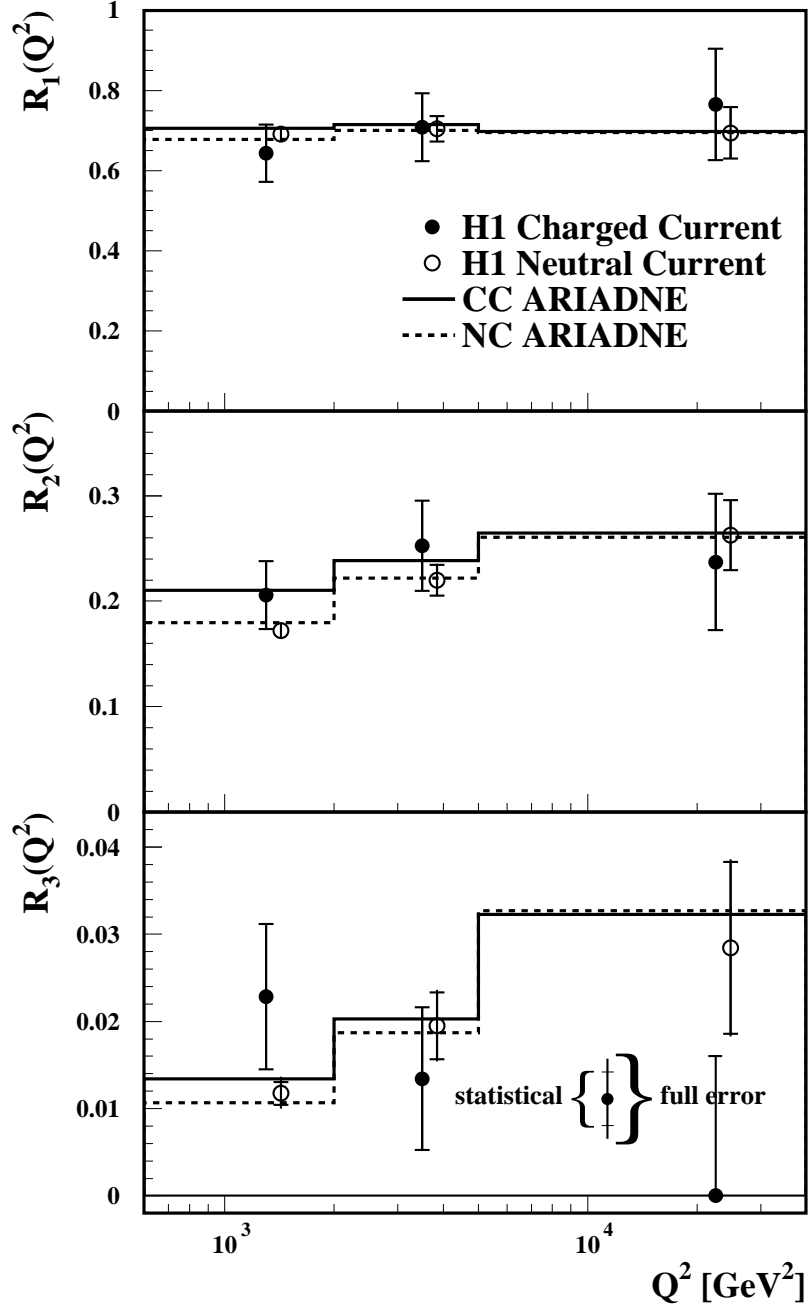,width=12.cm}
\end{center}
\caption{Rates of CC and NC events with one, two and three jets as a 
function of $Q^2$.  The events satisfy $p_T^{lept}>25$ GeV and $0.03 <
y < 0.85$.  The jets are reconstructed using the modified Durham
algorithm with a fixed jet resolution parameter $y_{cut} = 0.002$. 
The jets satisfy the cut $10^{\circ}<\theta_{jet} < 140^{\circ}$.
Also shown are the 
predictions of the MC model ARIADNE 4.10.
} 
\label{corr.jetrates.CCNC}
\end{figure}

The rates of events with one, two and three jets  $R_i(Q^2)$ are shown in 
Figure \ref{corr.jetrates.CCNC} as a function of $Q^2$ and are listed in 
Table \ref{tab.rates}. 
The jet event rates for CC and NC events are similar. For the chosen jet 
resolution, the fraction of one-jet events is $\approx$ 70\% for both CC or 
NC events. The fraction of dijet events is $\approx$ 20\% and that of 
three-jet events is a few percent. No CC three-jet events are yet observed at 
$Q^2 > 5000$ GeV$^2$, which is statistically consistent with the QCD model 
expectations. The 
$Q^2$ dependence of the jet event rates is small. Note that $R_1$ has 
a weak $Q^2$ dependence since most of the DIS events 
are reconstructed as (1+1) jet events. The jet event rates are well described 
by the QCD model ARIADNE.

\subsection{Differential dijet distributions}
\begin{figure}[t!]
\vspace*{-4.cm}
\begin{center}
\epsfig{file=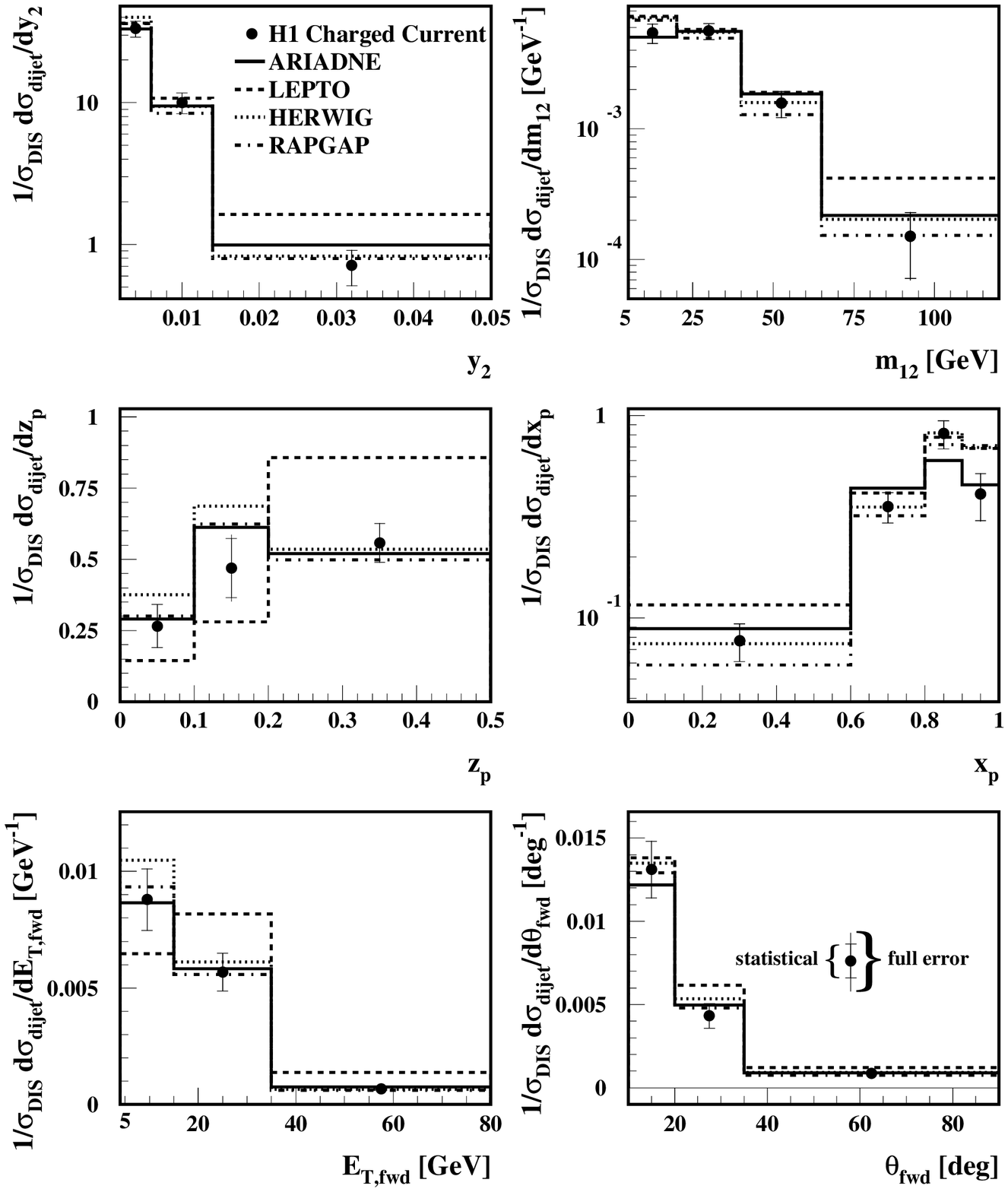,width=18.cm}
\end{center}
\vspace{-.5cm}
\caption{Distributions of $y_2$,  $m_{12}$, 
$z_p$, $x_p$, $E_{T, \; fwd}$ and $\theta_{fwd}$ in CC 
events with $p_T^{lept}>25$ GeV and $0.03 < y < 0.85$ determined with the 
modified Durham algorithm. 
The events satisfy the cuts $y_2 > 0.002$ and $10^{\circ}<\theta_{jet} < 140^{\circ}$. 
Also shown are the 
predictions of the MC models ARIADNE 4.10 (full line), LEPTO 6.5.2$\beta$ 
(dashed), HERWIG 5.9 (dotted) and RAPGAP 2.08/06 (dashed-dotted).}
\label{corr.jet.CC}
\end{figure}

The measured CC dijet distributions of $y_2$, $m_{12}$, $z_p$, $x_p$, 
$E_{T,\;fwd}$ and $\theta_{fwd}$, corrected for detector effects and 
the effects of QED radiation, are shown in Figure~\ref{corr.jet.CC}. 
The distributions are based on the 120 CC events that pass the requirements 
$y_2> 0.02$ and the $10^{\circ} < \theta_{fwd}< 140^{\circ}$. The measured 
differential 
dijet cross sections e.g. $d \sigma_{dijet}/dy_2$ are normalized by the 
inclusive 
DIS cross section $\sigma_{DIS}$ for the kinematic selection of section 2. 
The $y_2$ and $m_{12}$ distributions are steeply falling. In the tails of 
these 
distributions events with clear dijet structures (see Figure \ref{events}) 
and with dijet masses up to $\approx$ 100~GeV are observed.
The $z_p$ distribution shows a drop in the first bin at small $z_p$ which 
is due to the jet selection cut. The $x_p$ distribution is strongly peaked 
at large values of $x_p$ because the minimum $Q^2$ of the event selection is 
large. The corresponding average value of $\xi$ is $\approx$ 0.1.  
The forward jet distributions are strongly increasing at small polar angles 
and small transverse energies as is qualitatively expected by gluon 
bremsstrahlung off an incoming quark.

\begin{figure}[t!]
\vspace*{-4.cm}
\begin{center}
\epsfig{file=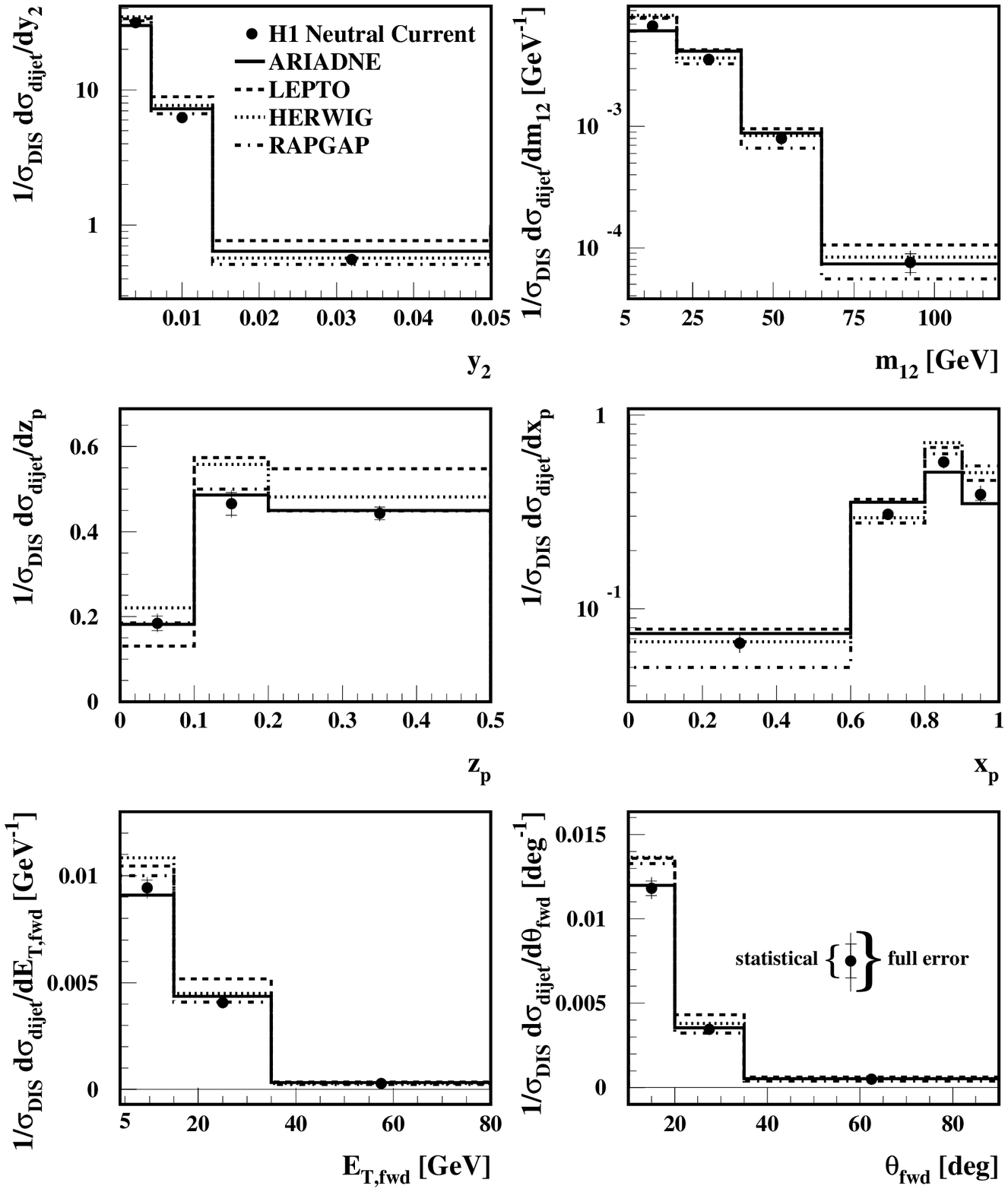,width=18.cm}
\end{center}
\vspace{-.5cm}
\caption{Distributions of $y_2$, $m_{12}$, 
$z_p$, $x_p$, $E_{T, \; fwd}$ and $\theta_{fwd}$ in NC events with 
$p_T^{lept}>25$ GeV and $0.03 < y < 0.85$ determined with the modified
Durham algorithm.  The events satisfy the cuts $y_2 > 0.002$ and $10^{\circ}
<\theta_{jet} < 140^{\circ}$.
Also shown are the 
predictions of the MC models ARIADNE 4.10 (full line), LEPTO 6.5.2$\beta$ 
(dashed), HERWIG 5.9 (dotted) and RAPGAP 2.08/06 (dashed-dotted).}
\label{corr.jet.NC}
\end{figure}

In Figure~\ref{corr.jet.NC} the corresponding distributions are shown for NC 
events. Here $\approx$ 1900 events remain after the jet selection. The NC 
distributions show the same features as the CC 
distributions. Note that due to the reduced statistical error of the NC 
distributions their total error is much smaller than that of the CC 
distributions. The differential CC and NC dijet distributions presented here 
are listed in Tables \ref{tab.nlo}. 

The corrected jet distributions are compared with the QCD 
models ARIADNE, HERWIG, LEPTO and RAPGAP. Within the errors, the 
data are reasonably well described by the QCD models ARIADNE, HERWIG and 
RAPGAP. LEPTO roughly follows the data distributions but overall it is 
inferior to the other models. Significant deviations from the data are 
observed in the $z_p$ distribution in particular. These observations are 
valid for both CC and NC distributions.

\subsection{Comparison with perturbative QCD calculations in NLO}

The differential dijet distributions are also compared with QCD predictions 
in NLO. 
Two sources of theoretical error on the QCD predictions have been considered: 
the uncertainty of the hadronization corrections and the renormalization 
scale uncertainties. The size of these uncertainties is similar for CC and NC 
events. The hadronization corrections are typically smaller than 10\%. 
Their uncertainty is estimated by the spread of the predictions of ARIADNE 
and HERWIG. The renormalization scale uncertainty of the NLO prediction is 
estimated by varying the renormalization scale $\mu_R^2$ from $Q^2$ to 1/4 
$Q^2$ and 4 $Q^2$. The resulting uncertainty is $\approx 5\%$. 
The non-remnant jets' average transverse energy in the Breit frame 
$\langle E_T^{Breit}\rangle $ is $\approx 10$ GeV. Choosing 
$\langle E_T^{Breit}\rangle^2$ 
as renormalization scale changes the NLO predictions by $\approx 5\%$ 
compared with the scale $Q^2$. A variation of the factorization scale has a 
marginal effect. 

\begin{figure}[t]
\vspace*{-4.cm}
\begin{center}
\epsfig{file=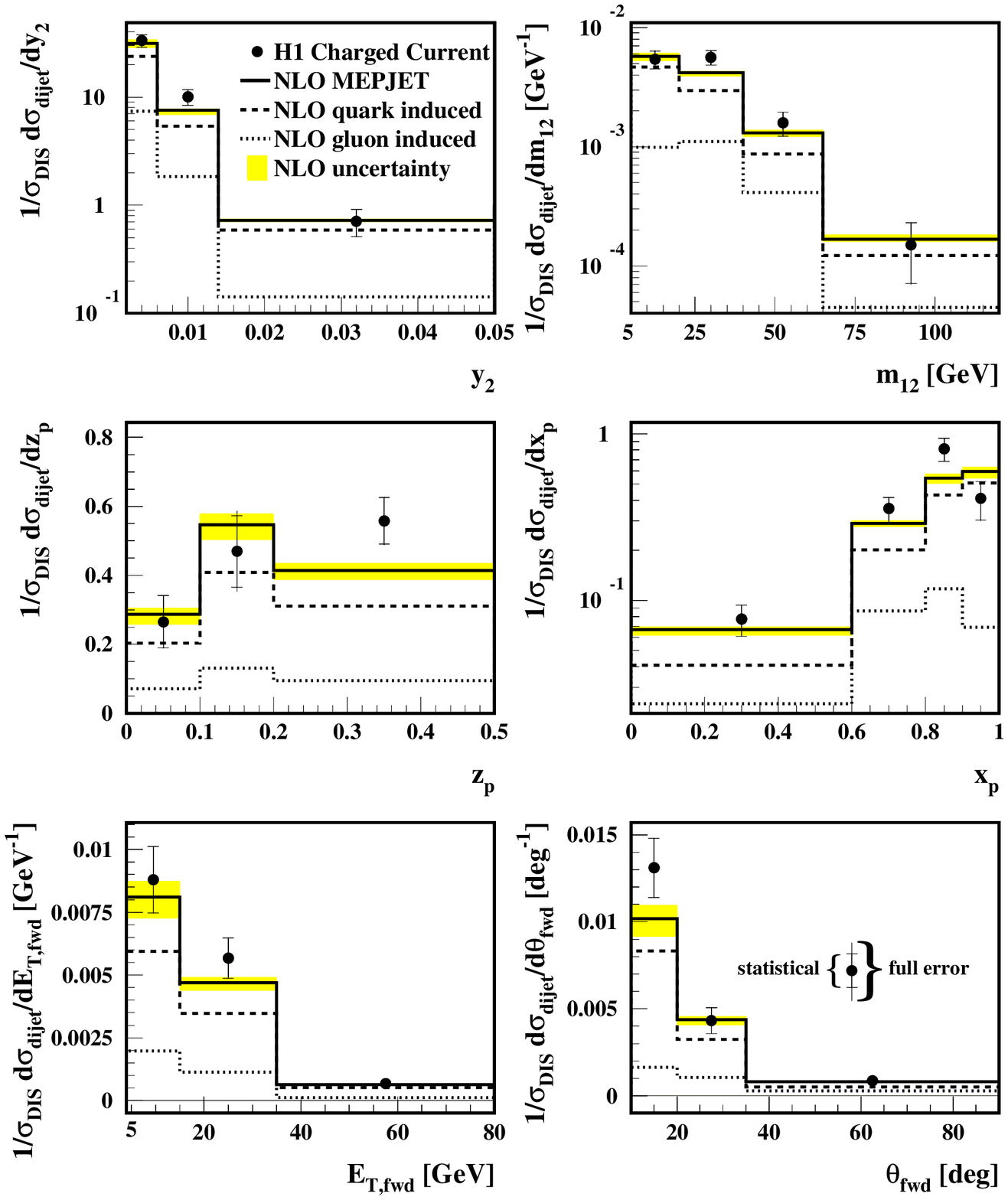,width=17.cm}
\end{center}\vspace*{-.5cm}
\caption{Distributions of $y_2$,  $m_{12}$, $z_p$, $x_p$, $E_{T, \; fwd}$ and 
$\theta_{fwd}$ in CC events with $p_T^{lept}>25$ GeV and $0.03 < y < 0.85$ 
determined with the modified Durham algorithm. The events satisfy the 
cuts $y_2 > 0.002$ and $10^{\circ}<\theta_{jet} < 140^{\circ}$. 
Also shown are perturbative 
QCD calculations in NLO obtained with MEPJET combined with a correction 
for hadronization effects. The shaded area shows the hadronization uncertainties and the renormalization scale uncertainties of the NLO calculations added 
in quadrature. In addition, the jet distributions obtained for quark-   
and gluon-induced processes are shown separately.}
\label{corr.jet.CC.NLO}
\end{figure}

\begin{figure}[t]
\vspace*{-4.cm}
\begin{center}
\epsfig{file=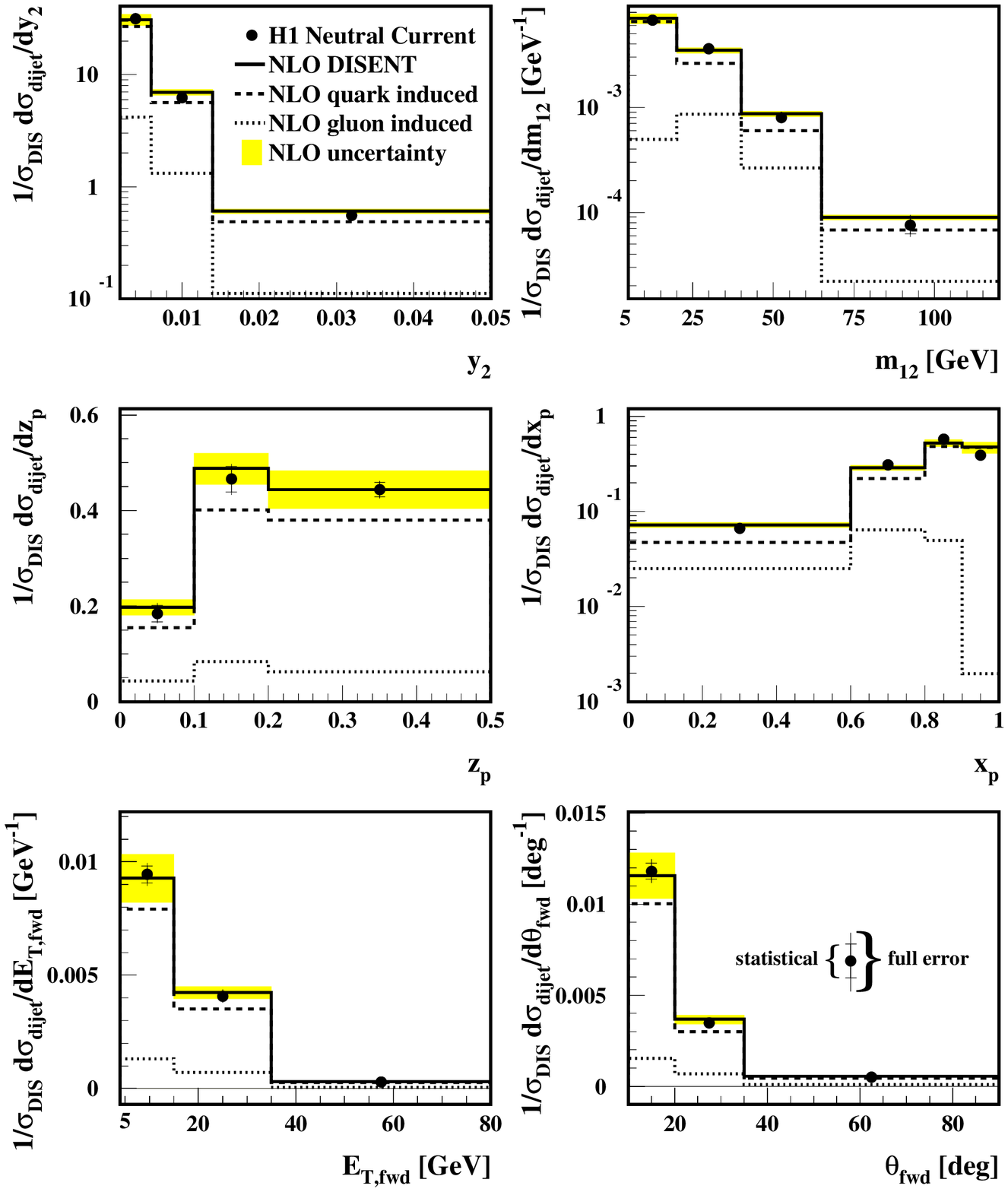,width=17.cm}
\end{center}\vspace*{-.5cm}
\caption{Distributions of $y_2$,  $m_{12}$, $z_p$, $x_p$, $E_{T, \; fwd}$ and 
$\theta_{fwd}$ in NC events with $p_T^{lept}>25$ GeV and $0.03 < y < 0.85$ 
determined with the modified Durham algorithm. The events satisfy the 
cuts $y_2 > 0.002$ and $10^{\circ} <\theta_{jet} < 140^{\circ}$. 
Also shown are perturbative 
QCD calculations in NLO obtained with DISENT combined with a correction 
for hadronization effects. The shaded area shows the hadronization uncertainties and the renormalization scale uncertainties of the NLO calculations added 
in quadrature. In addition, the jet distributions obtained for quark- and 
gluon-induced processes are shown separately.}
\label{corr.jet.NC.NLO}
\end{figure}

The CC jet distributions are compared with the NLO calculations of MEPJET 
in Figure~\ref{corr.jet.CC.NLO}. 
The MEPJET predictions provide a reasonable description of the data within 
errors. 
The corresponding NC distributions shown in 
Figure~\ref{corr.jet.NC.NLO} are well described by the NLO predictions of 
DISENT. 

In the Figures~\ref{corr.jet.CC.NLO} and \ref{corr.jet.NC.NLO} the 
NLO predictions for quark- and gluon-induced processes are also shown 
separately\footnote{Note that the normalizations $\sigma_{DIS}$ of the 
quark-induced, 
gluon-induced and combined dijet cross sections are identical.}.
The predicted fraction of gluon-induced dijet events is $\approx$ 20\% in 
CC and $\approx$ 15\% in NC scattering for the selection criteria of this 
analysis. These fractions change by less than one per cent when varying the 
factorization or renormalization scale in the QCD calculations from $Q^2$ to 
1/4 and 4 $Q^2$. The dominance of quark-induced processes is mostly due to 
the relatively large values of $x$ covered in this analysis. 
Figure \ref{corr.jet.CC.NLO} suggests that both quark and gluon 
contributions -- calculated using the parton density functions determined 
from inclusive measurements -- are needed to give a consistent description 
of dijet production in CC processes. 

\begin{figure}[t]
\vspace*{-3.cm}
\begin{center}
\epsfig{file=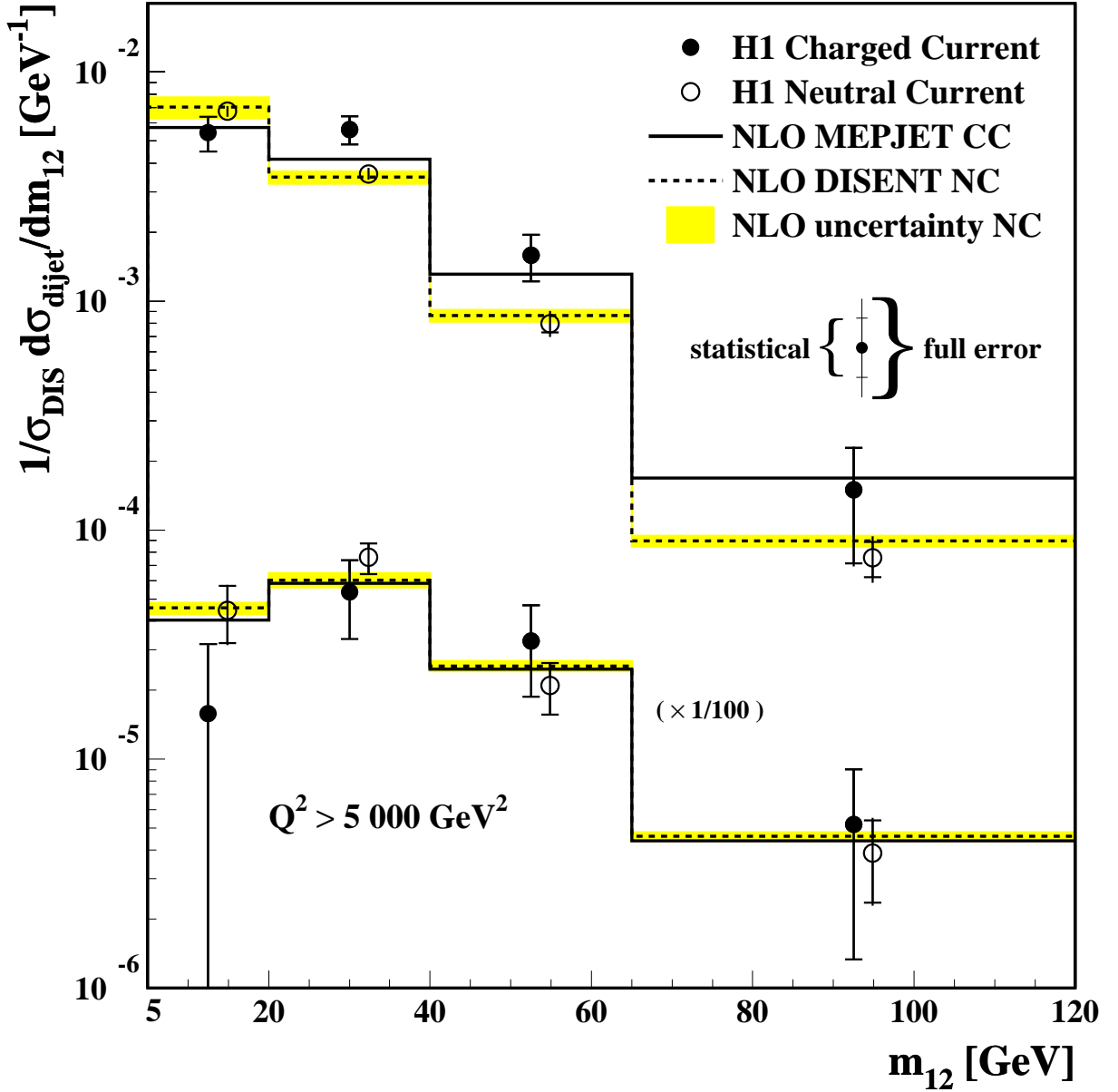, width=17.cm}
\end{center}
\vspace*{0.5cm}
\caption{Distributions of $m_{12}$ in CC events (full circles) and 
NC events (empty circles) for \mbox{$p_T^{lept}>25$~GeV} and $0.03 < y
< 0.85$ with and without the additional requirement $Q^2
>5000$~GeV$^2$ determined with the modified Durham algorithm.  The
events satisfy the cuts $y_2 > 0.002$ and $10^{\circ}
<\theta_{jet}~<~140^{\circ}$.
Also shown are perturbative QCD calculations in NLO obtained with MEPJET 
(for CC) and DISENT (for NC) 
combined with a correction for hadronization effects. The lower histograms and 
the corresponding data points have been scaled by a factor of 1/100. 
The shaded area shows the hadronization uncertainties and the 
renormalization uncertainties of the NLO calculations added in quadrature.
For clarity, the NC uncertainties are shown only. The CC uncertainties are 
of similar size.}
\label{corr.jet.CCNC.NLO}
\end{figure}

The CC and NC jet distributions have also been determined for the subsample 
of the selected DIS events with $Q^2 >$ 5\,000 GeV$^2$.
After the jet cuts 17 CC and 91 NC dijet events remain. In 
Figure~\ref{corr.jet.CCNC.NLO} the corresponding dijet mass distributions 
are compared with the NLO predictions of MEPJET and DISENT. Agreement is 
found in all bins. The 
measured distributions are also listed in Table \ref{tab.nlo5k}. 

\subsection{Direct comparison of CC and NC dijet distributions}

In Figure~\ref{corr.jet.CC.NCrew} the jet distributions of the CC events 
(full circles) are compared with those of the NC events selected in the 
same kinematic range (histogram). Systematic differences between the jet 
distributions are observed in several bins. This is expected due to the 
different electroweak 
couplings and gauge boson propagators which also lead to different kinematic distributions 
(see Figure \ref{kin.CC.NC.raw}). 
In order to account for these effects and to make possible a direct, model 
independent comparison of jet production in CC and NC processes, 
the NC events are reweighted 
to match the $x$ and $Q^2$ spectra of the CC events. 
The weights are given by the ratio of the inclusive NC and CC DIS cross 
sections at the $x$ and $Q^2$ of the NC event considered. The cross sections 
are calculated with DJANGO, and QED radiative corrections are taken 
into account. 
Note that the inclusive cross sections do not depend on the hadronic final 
state properties and thus the reweighting procedure is independent of the 
modelling of the hadronic final state. 

The effect of this procedure has been tested with the QCD models ARIADNE, 
HERWIG and RAPGAP, and the jet distributions of CC and reweighted 
NC events are predicted to agree within a few per cent typically. 
Residual differences between CC and reweighted NC jet distributions are 
expected due to the different fraction of gluon induced events in 
CC and NC processes, to helicity effects \cite{koerner,seymour}, and to the 
different parton densities contributing to NC and CC scattering. 
The NC event selection was repeated with the cut $P_T^{had}>25$~GeV instead 
of $p_T^{e}>25$~GeV. The changes in the corrected jet distributions are of 
the order of 2\%.

The measured NC jet distributions after reweighting are also shown in 
Figure~\ref{corr.jet.CC.NCrew}. They are found to be consistent with 
the CC distributions confirming that at short distances the QCD dynamics of 
the hadronic final state are essentially independent 
of the underlying electroweak scattering process as is expected within the 
standard model.

\section{Summary}

A sample of 460 CC and of $\approx$ 8\,600 NC events produced in 
deep-inelastic $e^+p$ scattering at HERA with the boson virtuality ranging 
from approximately $640 < Q^2 < 35\,000$~GeV$^2$
has been selected.  In this sample jets are reconstructed using a
modified version of the Durham algorithm.  Jet studies are hence extended into
a kinematic region where charged and neutral gauge bosons contribute
at comparable level.

Events with dijet structures are observed in CC processes. Differential CC 
dijet distributions are measured for the first time. Perturbative QCD 
calculations in NLO based on the 
MEPJET program describe the data well within errors. These calculations 
suggest that both quark and gluon contributions are needed to give a 
consistent description of dijet production in CC processes. 
The NC dijet distributions, measured in the same kinematic range, are well 
described by perturbative QCD predictions in NLO based on DISENT. 

The measured data sample contains events up to 
$Q^2 \approx 35\,000$ GeV$^2$ and $m_{12} \approx 100$ GeV and therefore 
probes QCD down to shortest distances.  Using parton densities derived
from NLO QCD fits to inclusive DIS cross sections, perturbative
calculations based on the electroweak and the strong ($O(\alpha_s^2)$)
matrix elements are found to give a consistent description of both the
NC and CC jet cross sections at highest dijet masses and highest
$Q^2$.

Comparison of the CC with the NC jet distributions confirms that at short 
distances the QCD dynamics of the 
hadronic final state are essentially independent of the underlying 
electroweak scattering process as is expected within the standard model.

\section*{Acknowledgements} 

We are very grateful to the HERA machine group whose outstanding efforts 
made this experiment possible. We acknowledge the support of the DESY 
technical staff. We appreciate the big effort of the engineers and 
technicians who constructed and maintain the detector. We thank the funding 
agencies for financial support of this experiment. We wish to thank the 
DESY directorate for the support and hospitality extended to the non--DESY 
members of the collaboration.

\clearpage\newpage

 \begin{table}[h!]
 \begin{center}
 \begin{tabular}{|c||c|r|r||c|r|r|  }
 \hline
  & \multicolumn{  3}{|c||}{CC} & \multicolumn{ 3}{|c|}{NC}\\
 \hline
 ${Q^2}$ [GeV]$^2$ & $R_1(Q^2)$ & $\delta_{stat}$ (\%)& 
$\delta_{sys}$(\%)&$R_1(Q^2)$ & $\delta_{stat}$(\%)&$\delta_{sys}$(\%)
 \\ 
 \hline
$600-2000$ & 0.64 & $\pm$11 & $\pm 1$
& 0.69 & $\pm$2.2 & $\pm 1.0$ \\
 \hline
$2000-5000$ & 0.71 & $\pm$12 & $\pm 1$
&  0.71 & $\pm$4.5 & $\pm 0.9$
 \\
 \hline
$>5000$ & 0.76 & $\pm$18& $\pm 1$
&  0.70 & $\pm$9.3 & $\pm 1.1$
 \\
 \hline\hline
 ${Q^2}$ [GeV]$^2$ & $R_2(Q^2)$ & $\delta_{stat}$(\%) & 
$\delta_{sys}$(\%)&$R_2(Q^2)$ & $\delta_{stat}$(\%)&$\delta_{sys}$(\%)
 \\ 
 \hline
$600-2000$ & 0.21 & $\pm$16 & $\pm 1$
& 0.172 & $\pm$3.5 & $\pm 2.5$
 \\
 \hline
$2000-5000$ & 0.25 & $\pm$17 & $\pm 2$ 
&  0.22 & $\pm$6.7 & $\pm 2.0$
 \\
 \hline
$>5000$ & 0.24 & $\pm$27 & $\pm 2$
&  0.26 & $\pm$13 & $\pm 2$
 \\
 \hline\hline
 ${Q^2}$ [GeV]$^2$ & $R_3(Q^2)$ & $\delta_{stat}$(\%) & 
$\delta_{sys}$(\%)&$R_3(Q^2)$ & $\delta_{stat}$(\%)&$\delta_{sys}$(\%)
 \\ 
 \hline
$600-2000$ & 0.022 & $\pm$37 & $\pm 5$
& 0.012 & $\pm$11 & $\pm 11$
 \\
 \hline
$2000-5000$ & 0.013 & $\pm$61& $\pm 6$ 
&  0.02 & $\pm$20& $\pm 8$
 \\
 \hline
$>5000$ & 0.0 & $+0.016$  & $-$
&  0.028 & $\pm$35 & $\pm 9$ \\
& & {\small (68\% CL)} & & & &
 \\
 \hline
 \end{tabular}
 \end{center}
\caption{Rates of CC and NC events with one, two and three jets as a 
function of $Q^2$.  The events satisfy $p_T^{lept}>25$ GeV and $0.03 <
y < 0.85$.  The jets are reconstructed using the modified Durham
algorithm with a fixed jet resolution parameter $y_{cut} = 0.002$. 
The jets satisfy the cut $10^{\circ}<\theta_{jet} < 140^{\circ}$.  The
relative statistical errors $\delta_{stat}$ and relative systematic
errors $\delta_{sys}$ are given in per cent.}
\label{tab.rates}
 \end{table}

\clearpage\newpage

 \begin{table}[h!]
 \begin{center}
 \begin{tabular}{|c||c|r|r||c|r|r|  }
 \hline
  & \multicolumn{  3}{|c||}{CC} & \multicolumn{ 3}{|c|}{NC}\\
 \hline
 $\mathbf{y_{2}}$    
&$\frac{\dps 1}{\dps \sigma_{DIS}}\frac{\dps {\rm d}\sigma_{dijet}}{\dps {\rm d}y_{2}}$ 
&$\delta_{stat}$(\%)&$\delta_{sys}$(\%)
&$\frac{\dps 1}{\dps \sigma_{DIS}}\frac{\dps {\rm d}\sigma_{dijet}}{\dps {\rm d}y_{2}}$ 
&$\delta_{stat}$(\%)&$\delta_{sys}$(\%)
\\ 
 \hline
$0.002 - 0.006$& 33
&$\pm$13&$\pm 2$
&31.4
&$\pm$3.5&$\pm 3.1$
 \\
 \hline
$0.006 - 0.014$& 10
&$\pm$17&$\pm 2$
& 6.3
&$\pm$5.4&$\pm 2.8$
 \\
 \hline
$0.014 - 0.05$& 0.7
&$\pm$28&$\pm 3$
&  0.56
&$\pm$8.9&$\pm 6.1$
 \\ \hline\hline
 $\mathbf{m_{12}}$ 
&$\frac{\dps 1}{\dps \sigma_{DIS}}\frac{\dps {\rm d}\sigma_{dijet}}{\dps {\rm d}m_{12}}$
&$\delta_{stat}$(\%)&$\delta_{sys}$(\%)
&$\frac{\dps 1}{\dps \sigma_{DIS}}\frac{\dps {\rm d}\sigma_{dijet}}{\dps {\rm d}m_{12}}$
&$\delta_{stat}$(\%)&$\delta_{sys}$(\%) \\ 
{\small [GeV]}& {\small[GeV$^{-1}$]} & & & {\small [GeV$^{-1}$]} & & \\ 
\hline
$5 -20$&  0.005
&$\pm$17&$\pm 4$
&  0.0067
&$\pm$4.0&$\pm 3.9$
 \\
 \hline
$20 - 40$&  0.006
&$\pm$14&$\pm 2$
&  0.0036
&$\pm$4.5&$\pm 3.9$
 \\
 \hline
$40 - 65$ &  0.0016
&$\pm$23&$\pm 5$
&  0.0008
&$\pm$9&$\pm 10$
 \\
 \hline
$65 - 120$&  0.0002
&$\pm$53&$\pm 12$
&  0.00008
&$\pm$18&$\pm 15$
 \\
\hline\hline
 $\mathbf{z_{p}}$ 
&$\frac{\dps 1}{\dps \sigma_{DIS}}\frac{\dps {\rm d}\sigma_{dijet}}{\dps {\rm d}z_{p}}$
&$\delta_{stat}$(\%)&$\delta_{sys}$(\%)
&$\frac{\dps 1}{\dps \sigma_{DIS}}\frac{\dps {\rm d}\sigma_{dijet}}{\dps {\rm d}z_{p}}$
&$\delta_{stat}$(\%)&$\delta_{sys}$(\%)
\\ 
 \hline
$0. - 0.1$&  0.27
&$\pm$29&$\pm 5$
&  0.18
&$\pm$9.4&$\pm 7.0$
 \\
 \hline
$0.1 - 0.2$&  0.47
&$\pm$22&$\pm 11$
&  0.47
&$\pm$5.7&$\pm 4.1$
 \\
 \hline
$0.2 - 0.5 $&  0.56
&$\pm$12&$\pm 3$
&  0.44
&$\pm$3.4&$\pm 2.4$
 \\
 \hline\hline
 $\mathbf{x_{p}}$       
&$\frac{\dps 1}{\dps \sigma_{DIS}}\frac{\dps {\rm d}\sigma_{dijet}}{\dps {\rm d}x_{p}}$
&$\delta_{stat}$(\%)&$\delta_{sys}$(\%)
&$\frac{\dps 1}{\dps \sigma_{DIS}}\frac{\dps {\rm d}\sigma_{dijet}}{\dps {\rm d}x_{p}}$
&$\delta_{stat}$(\%)&$\delta_{sys}$(\%)
\\ 
 \hline
$0. - 0.6$&  0.077
&$\pm$21&$\pm 2$
&  0.066
&$\pm$6.2&$\pm 9.9$
 \\
 \hline
$0.6 -0.8$&  0.36
&$\pm$17&$\pm 3$
&  0.31
&$\pm$4.9&$\pm 3.1$
 \\
 \hline
$0.8 -0.9$ &  0.81
&$\pm$16&$\pm 3$
&  0.57
&$\pm$5.0&$\pm 4.7$
 \\
 \hline
$0.9 -1.0$&  0.41
&$\pm$26&$\pm 6$
&  0.39
&$\pm$ 6.6&$\pm 9.0$
 \\
 \hline\hline
 $\mathbf{E_{T,fwd}} $
 &$\frac{\dps 1}{\dps \sigma_{DIS}}\frac{\dps {\rm d}\sigma_{dijet}}{\dps {\rm d}E_{T,fwd}}$
 &$\delta_{stat}$(\%)&$\delta_{sys}$(\%)&
$\frac{\dps 1}{\dps \sigma_{DIS}}\frac{\dps {\rm d}\sigma_{dijet}}{\dps {\rm d}E_{T,fwd}}$
 &$\delta_{stat}$(\%)&$\delta_{sys}$(\%)
\\
{\small [GeV]}& \small{[GeV$^{-1}$]} & & & \small{[GeV$^{-1}$]} & & \\ 
\hline
$4 - 15$&  0.0088
&$\pm$15&$\pm 2$
&0.0094
&$\pm$3.9&$\pm 3.7$
 \\
 \hline
$15 - 35$&  0.0057
&$\pm$14&$\pm 2$
&0.0041
&$\pm$4.4&$\pm 5.7$
 \\
 \hline
$35 - 80 $&  0.00067
&$\pm$26&$\pm 7$
&  0.00029
&$\pm$10&$\pm 15$
 \\
 \hline\hline
 $\mathbf{\theta_{fwd}}$
&$\frac{\dps 1}{\dps \sigma_{DIS}}\frac{\dps {\rm d}\sigma_{dijet}}{\dps {\rm d}\theta_{fwd}}$
&$\delta_{stat}$(\%)&$\delta_{sys}$(\%)
&$\frac{\dps 1}{\dps \sigma_{DIS}}\frac{\dps {\rm d}\sigma_{dijet}}{\dps {\rm d}\theta_{fwd}}$
&$\delta_{stat}$(\%)&$\delta_{sys}$(\%)
\\ 
\small{ [deg]}& \small{[deg$^{-1}$]} & & & \small{[deg$^{-1}$]} & & \\ 
\hline
$10 - 20$&  0.013
&$\pm$13&$\pm 2$
&  0.0118
&$\pm$3.7&$\pm 4.2$
 \\
 \hline
$20 - 35$&  0.0043
&$\pm$17& $\pm 2$
&  0.0035
&$\pm$5.3&$\pm 1.7$
 \\
 \hline
$35 - 90$ &  0.00086
&$\pm$23&$\pm 3$
&  0.0005
&$\pm$7.7&$\pm 3.4$ 
 \\
 \hline
 \end{tabular}
 \end{center}
\caption{Normalized dijet cross sections as a function of 
 $y_2$,  $m_{12}$, $z_p$, $x_p$, $E_{T, \, fwd}$ and 
$\theta_{fwd}$ in CC and NC events with $p_T^{lept}>25$~GeV and 
$0.03 < y < 0.85$ determined with the modified Durham 
algorithm. The events satisfy the cuts $y_2 > 0.002$ and 
$10^{\circ} < \theta_{jet} < 140^{\circ}$. The relative statistical errors 
$\delta_{stat}$ and systematic errors $\delta_{sys}$ are given in per cent.}
\label{tab.nlo}
 \end{table}
\clearpage\newpage

 \begin{table}[h!]
 \begin{center}
 \begin{tabular}{|c||c|r|r||c|r|r|  }
 \hline
  & \multicolumn{  3}{|c||}{CC} & \multicolumn{ 3}{|c|}{NC}\\
 \hline
 $\mathbf{m_{12}}$ 
&$\frac{\dps 1}{\dps \sigma_{DIS}}\frac{\dps {\rm d}\sigma_{dijet}}{\dps {\rm d}m_{12}}$
&$\delta_{stat}$(\%)&$\delta_{sys}$(\%)
&$\frac{\dps 1}{\dps \sigma_{DIS}}\frac{\dps {\rm d}\sigma_{dijet}}{\dps {\rm d}m_{12}}$
&$\delta_{stat}$(\%)&$\delta_{sys}$(\%)
 \\ 
{\small [GeV]}& {\small[GeV$^{-1}$]} & & & {\small [GeV$^{-1}$]} & & \\ 
\hline
 \hline
$5 - 20$ & 0.0016&$\pm$102&$\pm$11
&  0.0045&$\pm$28&$\pm 8$
 \\
 \hline
$20 - 40$ & 0.0054&$\pm$38&$\pm$4 
&  0.0076&$\pm$15&$\pm 2$
 \\
 \hline
$40 - 65$&   0.0033&$\pm$43&$\pm$4
&  0.0021&$\pm$26&$\pm 2$
 \\
 \hline
$65 -120$&  0.00052&$\pm$74&$\pm$7
&  0.00039&$\pm$39&$\pm 14$
 \\
 \hline
 \end{tabular}
 \end{center}
\caption{Normalized dijet cross sections as a function of  $m_{12}$ in CC and 
NC events with $Q^2>5000$~GeV$^2$, $p_T^{lept}>25$~GeV and $0.03 < y < 0.85$ 
determined with the modified Durham algorithm. The events satisfy the cuts 
$y_2 > 0.002$ and $10^{\circ} < \theta_{jet} < 140^{\circ}$. The relative statistical 
errors $\delta_{stat}$ and systematic errors $\delta_{sys}$ are given in 
per cent.}
\label{tab.nlo5k}
 \end{table}

\setlength{\unitlength}{1 mm}
\clearpage
\newpage

\newpage
\clearpage

\begin{figure}[h!]
\begin{center}
\begin{picture}(150.,150.)
\put(-14,-65)
{\epsfig{file=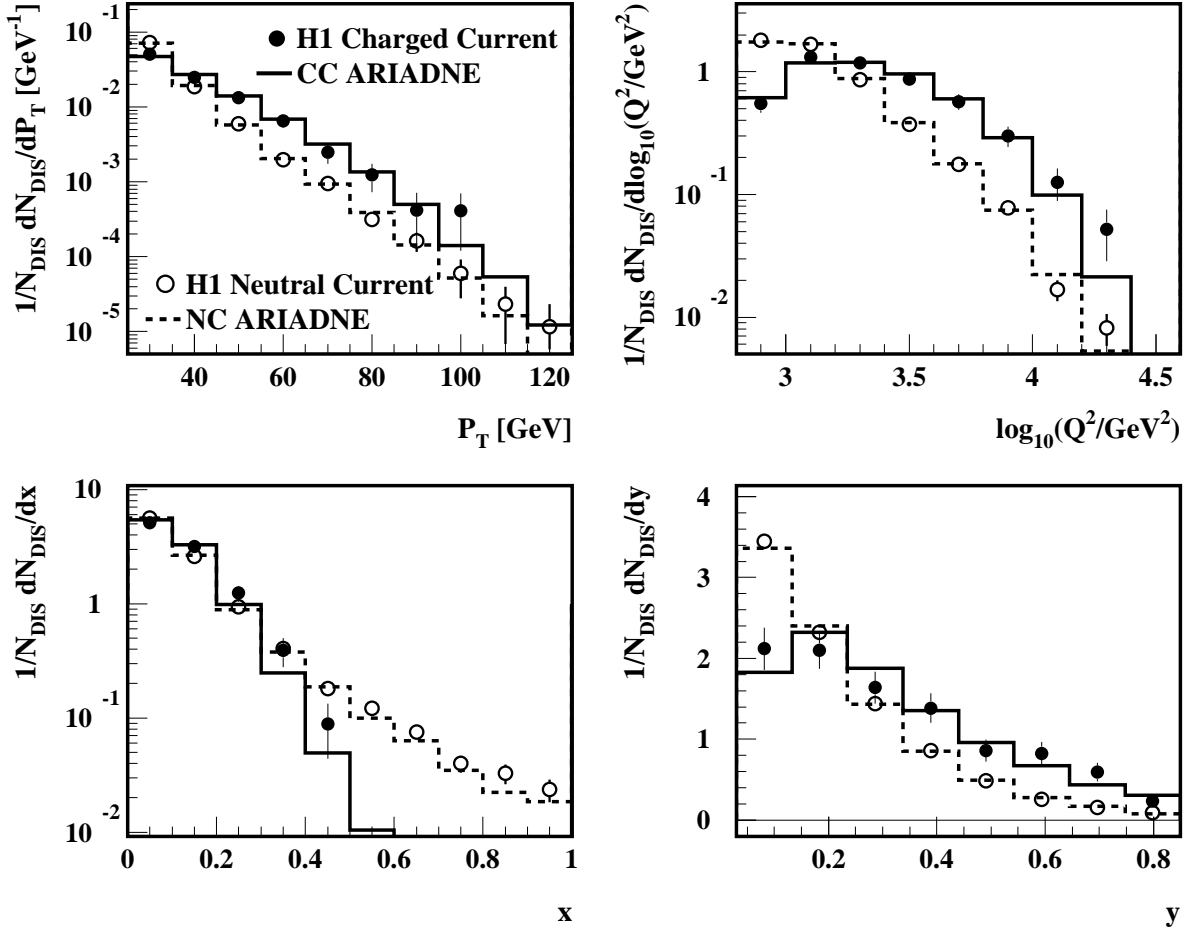,width=17.5cm}}
\end{picture}
\end{center}
\caption{Uncorrected data distributions of $P_T$, $Q^2$, $x$ and $y$ 
for the selected CC and NC events. The observables are calculated using the 
hadronic final state for CC events, and the scattered positron for NC events. 
The errors are statistical only. Also shown are the 
corresponding predictions of the MC model ARIADNE 4.10 including 
radiative QED corrections and the H1 detector simulation for CC processes 
(full line) and NC processes (dashed).}
\label{kin.CC.NC.raw}
\end{figure}

\begin{figure}[t]
\begin{center}
\begin{picture}(150.,150.)
\put(0,-20){\epsfig{file=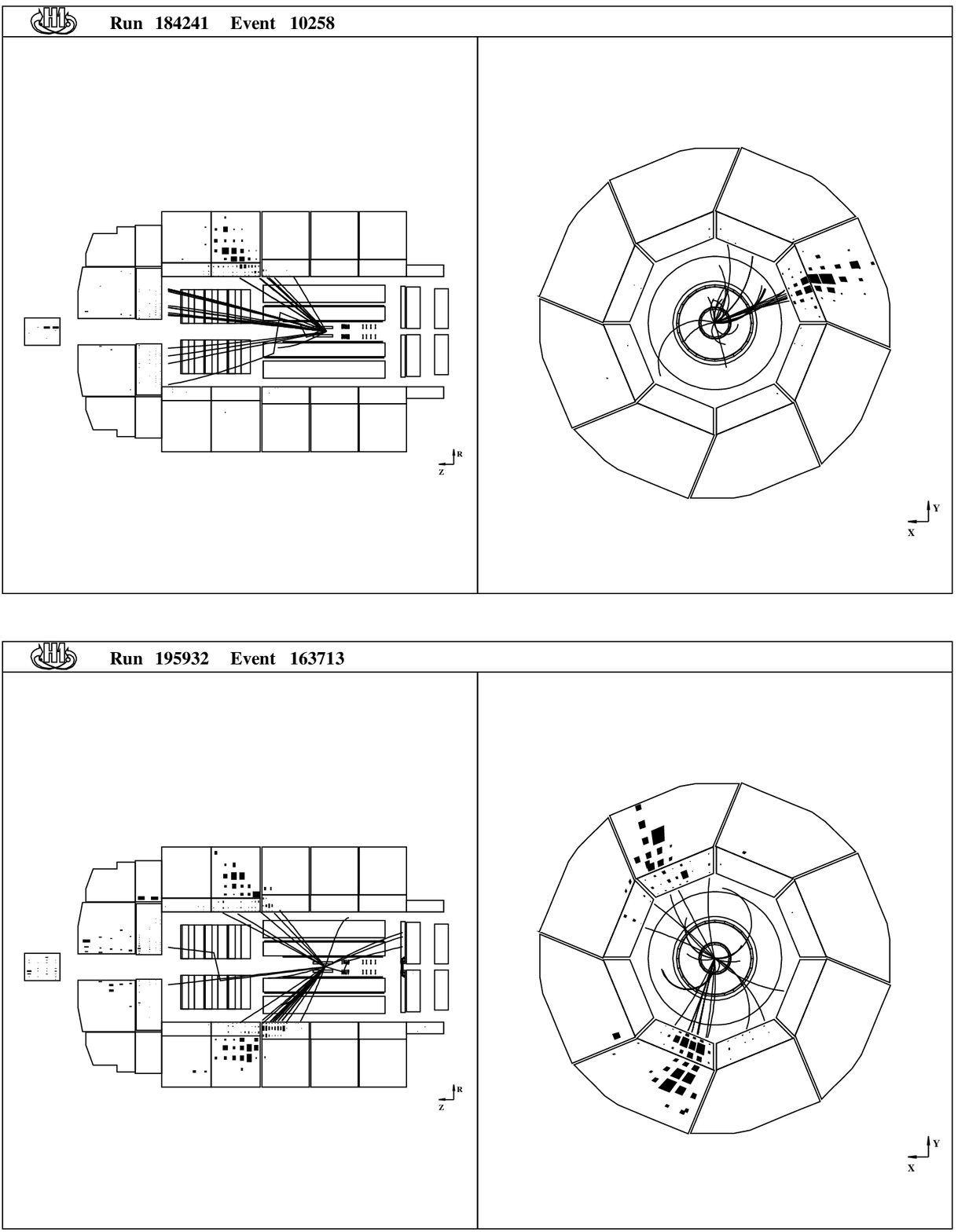,width=15.cm}}
\end{picture}
\end{center}
\vspace{2.0cm}
\caption{A display of two CC events. The left part 
shows a side view of the H1 central and forward tracking systems 
surrounded by the electromagnetic and hadronic sections of the 
liquid argon calorimeter and of the lead/scintillating-fibre 
calorimeter. The full lines and filled rectangles  correspond 
to tracks reconstructed in the tracking systems and energy depositions 
in the calorimeter, respectively. The proton beam enters from the right. 
The right part shows a view along the beam of the same events.
For the upper event 
$y_2 \approx 0.00008$ and 
$m_{12}\approx 12$~GeV. For the the lower event $y_2\approx 0.013$ and 
$m_{12}\approx 73$~GeV.
}
\label{events}
\end{figure}

\begin{figure}[t]
\vspace*{-2.cm}
\begin{center}
\begin{picture}(100.,250.)
\put(-30,25)
{\epsfig{file=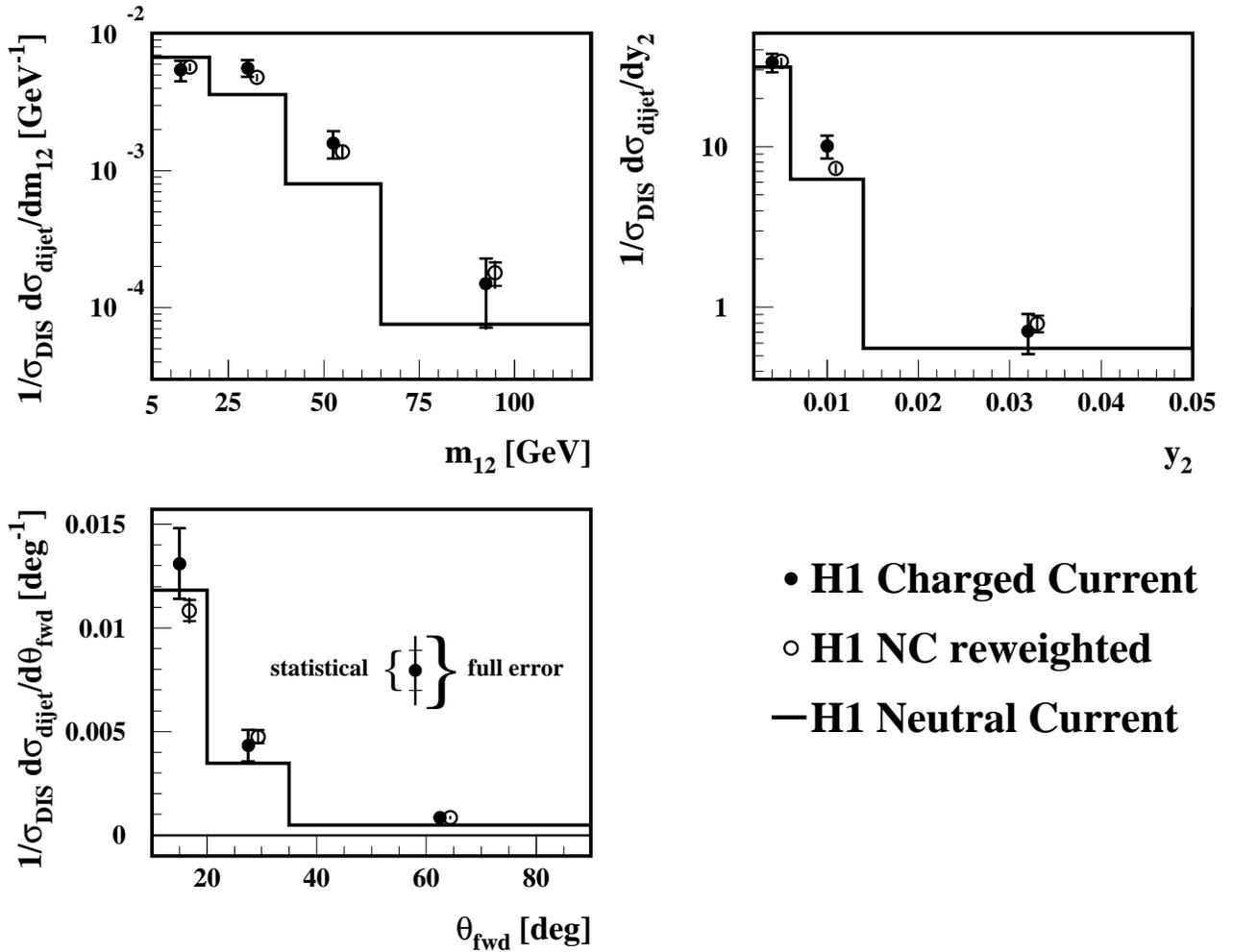, width=18.cm}}
\end{picture}
\end{center}
\vspace*{-4.5cm}
\caption{The distributions of  $m_{12}$, $y_2$ and $\theta_{fwd}$ in CC 
events (full circles) and the corresponding distribution in NC events with 
reweighting (empty circles) as described in the text. The solid histogram 
corresponds to the NC distributions without reweighting, which are also shown 
as data points in Figures \ref{corr.jet.NC} and \ref{corr.jet.NC.NLO}. The 
same jet selection criteria as above are applied. 
} 
\label{corr.jet.CC.NCrew}
\end{figure}

\end{document}